# Thermoelectric Properties of (1-x)$LaCoO_3$.x$La_{0.7}Sr_{0.3}MnO_3$ Composite

Ashutosh Kumar[a,$], Karuna Kumari[a], B. Jayachandran[b], D. Sivaprahsam[b] and Ajay D Thakur[a,*]

[a]*Department of Physics, Indian Institute of Technology Patna, Bihta-801 106, India*
[b]*ARC-International, Indian Institute of Technology Madras, Research Park, Chennai-600 133, India*

[*] *ajay.thakur@iitp.ac.in*, [$]*ashutosh.pph13@iitp.ac.in*

We report the thermoelectric (TE) properties of (1-x)$LaCoO_3$.x$La_{0.7}Sr_{0.3}MnO_3$ ($0 < x < 0.10$) composite in a temperature range 320-800 *K*. Addition of $La_{0.7}Sr_{0.3}MnO_3$ to $LaCoO_3$ in small amount (5 weight %) improves the overall Seebeck coefficient ($\alpha$) at higher temperatures. The electrical conductivity however decreases due to a decrease in carrier concentration of the composite. The decrease in electrical conductivity of the composite at high temperature may be attributed to the insulating nature of the LSMO above room temperature. Thermal conductivity ($\kappa$) of all the samples increases with an increase in the temperature, but decreases with increasing LSMO content. We also report the local variation of Seebeck coefficient across the composite samples measured using a precision Seebeck measurement system. A maximum value of 0.09 for the figure of merit (*ZT*) is obtained for 0.95$LaCoO_3$.0.05$La_{0.7}Sr_{0.3}MnO_3$ at 620 K which is significantly higher than the *ZT* of either of $LaCoO_3$ or $La_{0.7}Sr_{0.3}MnO_3$ at 620 K. This suggests the potential for enhancement of operating temperatures of *hitherto* well known low temperature thermoelectric materials through suitable compositing approach.

# 1. Introduction

Thermoelectric (TE) materials convert heat energy into electrical energy and vice versa *via* TE phenomena in semiconductors/semi-metals [1,2]. In particular, power generators capable of working in ambient air for long durations at intermediate temperature (300-800 *K*) are highly desirable for harvesting waste heat from vehicle engines and industrial chimneys. [3,4] For such demands, oxide ceramics are highly promising, but the corresponding TE conversion efficiencies are poor at present. The efficiency of a thermoelectric generator (TEG) is a function of TE figure of merit (ZT=$\alpha^2\sigma T/\kappa$, where α, σ, T and $\kappa$ are the Seebeck coefficient, electrical conductivity, absolute temperature and thermal conductivity respectively, $\alpha^2\sigma$ is called power factor); an important parameter which determines the utility of a material for use in TE generators. Nevertheless, TE materials have occupied only niche markets because of their low efficiency. An efficient TE material should have low $\kappa$, high $\alpha$ as well as a high $\sigma$. These TE parameters are dependent on each other and hence an improvement in one of the parameters often leads to undesirable changes in the other parameters, e.g. an enhancement in the electrical conductivity leads to a corresponding increase in the electronic part of thermal conductivity (in accordance with the Wiedemann-Franz law), and therefore it is a very challenging task to improve *ZT*.

Several theoretical studies predicted the possibility to increase TE properties in composites by either increasing power factor [5,6] or by reducing κ due to phonon scattering at the interfaces [7-8]. Kim *et al.* demonstrated that ZT value of TE materials can be enhanced *via* reduction of κ by embedding nanoparticles in crystalline semiconductor [9]. TE properties of several composites show the reduction of electrical resistivity and κ due to increase in charge carrier scattering and phonon scattering at the interface in composites [10-12]. Also, α can be enhanced by carrier filtering effect caused by the interfacial barrier in nanocomposites or even in bulk polycrystalline composite systems [13,14].

Among oxide TE materials, cobalt-based compounds are of interest due to the possibility of accessing different spin states [15]. The nature of Seebeck coefficient (α) in $LaCoO_3$ (LCO) is extremely sensitive to doping and temperature [16,17]. A large value of Seebeck coefficient ($\alpha$) occurs in LCO system, but with low electrical conductivity [18]. Androulakis *et al.* [19] showed highest *ZT* (~0.18) in $La_{0.95}Sr_{0.05}CoO_3$ at room temperature which is considerably high compared to other oxide systems. In LCO system, the conductivity at high temperature is very good but the Seebeck coefficient reduces significantly. However, improved TE properties were found in the LCO system by either substitution at La/Co site or by co-substitution [20], but, at a higher temperature (>500 K) the ZT value is very low in the substituted LCO system [21].

On the other hand, oxide composites have shown promising values of ZT at higher temperatures where the power factors were found to increase along with a decrease in thermal conductivity due to the aforementioned phenomenon [22-24]. Also, a large number of approaches have been followed to enhance the ZT values of oxide composite including reduction of average grain size or by chemical substitution [25,26]. Moreover, the improvement in electrical conductivity has been achieved at the cost of Seebeck coefficient and hence the overall ZT is not significantly enhanced at high temperature. To reach a high value of *ZT* at higher temperatures by adopting the above-mentioned strategies, we studied a series of LCO-based oxide composite by introducing $La_{0.7}Sr_{0.3}MnO_3$ (LSMO, a p-type perovskite with fascinating physical properties [27]). The crystal structure of the two materials is very similar. To the best of our knowledge, no systematic study has been done on TE properties of LCO/LSMO composite. In this report, we have studied the TE properties of $(1-x)LaCoO_3.xLa_{0.7}Sr_{0.3}MnO_3$ for $(0.00 \leq x \leq 0.50)$.

## 2. Experimental

Polycrystalline samples of $LaCoO_3$ and $La_{0.7}Sr_{0.3}MnO_3$ were synthesized using the standard solid-state method. A stoichiometric amount of $La_2O_3$, $SrCO_3$, and $Mn_2O_3$ precursor powders were weighed for the synthesis of LSMO; $La_2O_3$ and $Co_3O_4$ powders were weighed for the synthesis of LCO. All the precursors are of high purity (Sigma Aldrich, 99.99%). The weighed mixture of these precursors was mixed thoroughly in acetone. The dried mixture was kept in alumina crucibles and calcined at 1473 *K* for 12 hours in a muffle furnace with 4˚/min heating and cooling rate. The calcined powders were ground for 30 minutes to make it more homogenous. The synthesized LSMO and LCO powders were weighed and mixed at weight ratios satisfying (1-x)LCO.xLSMO, where, x=0.00, 0.05, 0.10, 0.15, 0.20, 0.50. LCO-LSMO mixture was again mixed thoroughly and consolidated into three sets of pellets of 20 mm diameter and was sintered at 1523 K for 12 hours with slow cooling and heating rate (2˚/min). Crystallographic structure and phase identification was done using X-ray diffraction (XRD) on a Rigaku diffractometer with Cu-Kα radiation ($\lambda=1.5406$ Å) with a scan rate of $1^{\circ}$/min and step size of $0.02^{\circ}$. Field emission scanning electron microscopy (FE-SEM, Zeiss) was performed for microstructural information. One of the pellets was cut into rectangular bars of 12*mm*×4*mm*×4*mm* dimension using a diamond cutter (IsoMet® Low Speed Saw) in the presence of escort oil (Buehler) and the same sample was ultra-sonicated for 30 minutes to remove any impurity during cutting of the samples. The rectangular bar-shaped sample was used to measure Seebeck coefficient and electrical resistivity using Seebsys™ (NorECs AS, Norway) under the conventional four-probe method in the temperature range 320 - 800 *K*. For measurement of $\alpha$, a temperature difference of 10 *K* was maintained between both ends of the sample using an auxiliary heater at one end of the sample. Precision Seebeck measurement (PSM PANCO, Germany) system was used to measure the local values of Seebeck coefficient across the sample. Electrical conductivity ($\sigma$) was observed by taking the inverse of the electrical resistivity data in the entire temperature range. Remaining two pellets of 20 *mm* diameter were used for thermal conductivity measurement using nonsteady state**,** transient plane source (TPS) technique which utilizes a sensor element, made of 10 $\mu m$ thick Nickel- metal in the shape of a double spiral. The sensor is sandwiched between the two pellets, in which room temperature thermal conductivity measurement was done by supplying 100 *mW* power for 10 seconds. The room temperature optimized value of parameters including laser power and measurement time was used to measure the high-temperature thermal conductivity of all the samples. The measurement errors for Seebeck coefficient, electrical conductivity, and thermal conductivity were about 3%, however, the corresponding error in the measurement of power factor could be up to about 10% [28-30].

## 3. Results and discussion

Figure 1 shows the XRD pattern of $(1-x)LaCoO_3.xLa_{0.7}Sr_{0.3}MnO_3$ ($0.00\leq x\leq 0.50$) composite sample calcined at 1473 K. For x=0.00, we observed diffraction peak due to LCO only with characteristic 2θ peaks at 32.88° and 33.20° and is in agreement with ICDD File no. 00-048-0123. It has been observed that the with addition of LSMO in $(1-x)LaCoO_3.xLa_{0.7}Sr_{0.3}MnO_3$ composite, we found the diffraction peak corresponding to LSMO (2θ=32.49°, 32.70°) and the intensity pertaining to LSMO become stronger with increasing weight percentage (x) of LSMO. Also, in the diffraction pattern of composites, only peaks corresponding to LCO and LSMO phases were observed and no secondary phase peaks were observed within the sensitivity of XRD, which suggests that LCO and LSMO phases are compatible and coexist without forming secondary phases. However, it is believed that interatomic diffusion between the LCO and LSMO phases can occur during sintering at 1523 K due to intersite diffusion between Mn and Co as well as between La and Sr because of their comparable ionic radii [31]. Rietveld refinement of all the composite samples was done using FullProf™ software to further investigate the lattice parameters and atomic position of the two phases in the composite with space group R-3c. Rietveld refinement parameters $R_{exp}$, $R_{wp}$, $R_p$, $\chi^2$, lattice parameters, the volume of the unit cell are provided in Table I, atomic positions with their occupancy are in Table II and the corresponding refinement pattern for $(1-x)LaCoO_3.xLa_{0.7}Sr_{0.3}MnO_3$ with x=0.05 is shown in Fig. 2.

Figures 3(a-d) present the typical FESEM images of the $(1-x)LaCoO_3.xLa_{0.7}Sr_{0.3}MnO_3$ ($0.00 \leq x \leq 0.50$) composite samples synthesized using the solid-state route. The grains are observed to be densely connected with a similar grain size of between 0.5 and 1 μm owing to a high-temperature sintering. It can be noted that no new phase was formed in the high-temperature sintering process and the phase identities of LCO and LSMO phases were preserved.

The high-temperature thermoelectric measurements of $(1-x)LaCoO_3.xLa_{0.7}Sr_{0.3}MnO_3$ ($0.00 \leq x \leq 0.50$) were done from 320-800 *K*. Temperature variation of Seebeck coefficient (α) for all the composite samples is plotted as shown in Fig. 4(a). All the samples display transitions from negative to positive values of α with an increase in temperature. The value of *α* for LCO is found to be negative which may be attributed to the oxygen deficiency in LCO system and is in agreement with the previous reports [32]. Koshibae *et al.* [33] demonstrated that the strong correlation of 3d electrons and the characteristic spin degeneracy in cobalt oxide causes the large value of Seebeck coefficient. The absolute values of α are found to decrease with x>0.00 in $(1-x)LaCoO_3.xLa_{0.7}Sr_{0.3}MnO_3$ composite. It is assumed that the electronic configuration of the Mn and Co elements play a crucial role in determining the thermoelectric response [31,34]. In the presents composite oxide system, Mn and Co ions exist in the form of $Mn^{3+}$, $Mn^{4+}$, and $Co^{3+}$, respectively. Based on Hund's rule and crystal field splitting, the electronic configuration of Mn and Co in LSMO and LCO are- $Mn^{3+}$:$t_{2g}^3e_g^1$, $Mn^{4+}$:$t_{2g}^3$ $e_g^0$ and $Co^{3+}$:$t_{2g}^6$ $e_g^0$, respectively among which $e_g^1$ electron of $Mn^{3+}$ is the only electronically active species [31, 34]. The interaction of Mn with Co ion ($Mn^{3+}$-O-$Co^{3+}$) leads to a decrease in the proportion of $Mn^{3+}/Mn^{4+}$ which results in the localization of $e_g^1$ electron of $Mn^{3+}$. It has also been observed that *α* increases from negative values to positive values up to 560

*K* and then decreases with increase in temperatures. This remarkable increase in *α* with an increase in temperature up to 550 *K* is attributed to the spin state transitions in the LCO system [35,36]. The value of *α* goes from negative to positive values and reaches a maximum value of ~ 120 *μV/K* at 500 *K* and then decreases with further increase in temperature for x=0.00. However, in cobaltates, it is very difficult to ascertain the exact nature of charge carriers from the sign of α alone [37,38]. For x=0.05, *α* shows a similar behavior with temperature (~ -102 *μV/K* at 320 K) and it increases towards positive Seebeck values and reaches a maximum value of Seebeck coefficient ~117 *μV/K* at 520 *K* and then decreases with further increase in temperature which is high compared to the co-substituted $LaCoO_3$ system [20,21]. It has been observed that with the addition of p-type LSMO sample in LCO, the values of Seebeck coefficient at a higher temperature (>480 *K*) decreases slowly with increase in temperature as compared to the LCO sample. This may be due to the increase in resistivity of the composite sample as compared to LCO sample, as the interdiffusion between Co and Mn ions lead to localization of electronically active carriers which leads to a high value of *α* at higher temperatures. This increase in resistivity of LSMO and increase in electrical conductivity of LCO at a higher temperature form a potential across the LCO/LSMO interface, which leads to increase in the value of α at higher temperature comparing to LCO system. It has been further observed that with the increase in the value of x in the composite sample, the value of *α* near room temperature decreases from -123 *μV/K* for x=0.00 to -75 *μV/K* for x=0.50. This may be attributed to the competition between p-type LSMO and the n-type LCO near room temperature. The inclusion of LSMO in LCO is found to be very useful in maintaining high Seebeck coefficient of the composite sample at higher temperatures (> 480 K).

Figure 4 (b) shows the variation of electrical conductivity (*σ*) for all the composite sample from 320-800 *K* measured using standard DC four-probe technique. It has been observed that at 320 *K* the conductivity is maximum for x=0.00 and decreases slightly with the increase in the value of x in the composite. This suggests that addition of LSMO in LCO enhances the interaction between different electronic states of $Co^{3+}$ and $Mn^{3+}$ leading to the localization of $e_g^1$ electrons and hence resistivity increases. It is also believed that electronic states in manganites are more localized compared to that of cobaltate [31]. It has been observed that all the sample shows an increase in conductivity with an increase in temperature. The maximum value of conductivity was found to be ≈59000 *S/m* for x=0.00 at 800 *K* and the value is identical for x=0.05 sample at the same temperature. However, with the increase in the value of x in the composite, the value of conductivity decreases compared to the parent system. The increase in electrical conductivity with an increase in temperature shows the semiconducting nature of the composite samples.

In the composite sample, the Seebeck coefficient is a combination of individual Seebeck coefficient due to two different phases present. In order to check the homogeneity of Seebeck coefficient observed in the composite samples, we have performed precision Seebeck measurement (PSM) across 0.5mm*0.5mm area of the composite samples. The PSM values observed for the composite samples are shown in Fig. 5(a-d). It has been observed that the values of α vary across the sample. The variation in Seebeck coefficient is less for x=0.05 (-88 $\mu V/K$ to -105 $\mu V/K$) where a major portion of the area has an α value of around -100 $\mu V/K$ as shown in Fig 5(a). The variation in α is found to increase with the increase in the value of x for the composite samples. This may be attributed to the increase in the number of grains of LSMO having a smaller Seebeck coefficient compared to LCO due to their electronic structure. For x=0.50 sample, we have observed local traces of positive values of Seebeck coefficient (~ 8 $\mu V/K$) which may due to the presence of LSMO polycrystalline sample.

Figure 6 (a) shows the total thermal conductivity ($k$) as a function of temperature from 320-800 $K$ for (1-x)$LaCoO_3$.x$La_{0.7}Sr_{0.3}MnO_3$ composite sample. Thermal conductivity ($\kappa$) consists of phonon thermal conductivity ($\kappa_{ph}$) and electronic thermal conductivity ($\kappa_e$), calculated using $\kappa_e=L_0\sigma T$, where, $L_0$ is the Lorentz number (1.50× $10^{-8}$ $V^2/K^2$) for non-degenerate semiconductors [39]. The $\kappa_{ph}$ can be calculated by subtracting the $\kappa_e$ part from total thermal conductivity $\kappa$. The electronic part of thermal conductivity calculated using measured electrical conductivity and Wiedemann-Franz law is shown in Fig. 6 (b) and phonon thermal conductivity ($\kappa_{ph}$) is shown in Fig 6 (c). Total thermal conductivity has been found to increase with an increase in temperature for (1-x)$LaCoO_3$ .x$La_{0.7}Sr_{0.3}MnO_3$ composite which may be due to a dramatic increase in the $\kappa_e$. This shows that thermal conductivity varies differently for x=0.00 compared to the composite samples. For LCO, thermal conductivity increases almost linearly up to 500 $K$ and above this temperature the value of $\kappa$ starts saturating, this behavior may be due to the fact that at high temperature where the atomic displacements are large, more number of phonons participate in the collision leading to a decrease in the phonon mean free path and hence reduces the thermal conductivity. The addition of LSMO into LCO reduces the thermal conductivity. As shown in Fig. 6 (a), $\kappa$ reduces from 2.67 $W/m\text{-}K$ for x=0.00 to 2.08 for x=0.50 at 800 $K$. This reduction in $\kappa$ may be due to a cumulative effect of the enhancement of phonon scattering and reduction in electronic conductivity with the increase in LSMO content at high temperature. It has been observed that total thermal conductivity $\kappa$ is dominated by phonon thermal conductivity $\kappa_{ph.}$ Although electronic thermal conductivity is enhanced due to high values of electrical conductivity, however phonon contribution of total thermal conductivity is 60 - 80 % of the total thermal conductivity, suggesting that *ZT* can be further enhanced by reducing the lattice part of thermal conductivity by either proper substitutions or optimized microstructure to enhance phonon boundary scattering.

Based on the values of α, σ, and κ, the figure of merit (*ZT*) is estimated for all the composite samples and is plotted as a function of temperature and shown in Fig. 7. A noticeable enhancement in the figure of merit has been observed for the x=0.05 composite sample. The figure of merit for x=0.05 increases from the room temperature and reaches a maximum at 620 *K* (*ZT*=0.09) which is around one order in magnitude larger than that of pure LCO (*ZT*~0.01) and larger than the other oxide systems at the same temperature [40-42]. As shown in Fig. 7, the maximum value of *ZT* for LCO is 0.03 at 560 *K*. This suggests that slight addition of LSMO (x=0.05 in (1-x)$LaCoO_3$ .x$La_{0.7}Sr_{0.3}MnO_3$) not only enhances the value of *ZT* but also enhances the operating temperature of the TE materials. For x=0.10, the maximum value of *ZT* ~0.047 at 640 *K* and with further increase in LSMO content the *ZT* values are found to decrease as the power factor decreases.

## 4. Conclusions

(1-x)$LaCoO_3$.x$La_{0.7}Sr_{0.3}MnO_3$ composite samples have been prepared using the standard solid-state method. Structural properties were studied extensively using XRD pattern followed by Rietveld analysis. Surface morphology and precise Seebeck measurements were done using FESEM and precision Seebeck measurement system respectively. From these measurements, we found the homogenous surface morphology and uniform Seebeck coefficient values across the samples. Precision Seebeck measurement of the composite samples shows the slight variation of Seebeck coefficient with the increase in LSMO content. Seebeck coefficient of (1-x)$LaCoO_3$ .x$La_{0.7}Sr_{0.3}MnO_3$ composite samples were first found to increase with the increase in temperature up to 550 *K* then decrease with the further increase in temperature. The increase in electrical conductivity was also observed with increasing temperature and its value reached ≈ 59000 *S/m* for x=0.05 at 800 *K*, indicating the metallic nature of the samples at high temperatures. The decrease in Seebeck coefficient and increase in electrical conductivity with a temperature of all the samples are strongly affected by the spin state transition. The total thermal conductivity of these oxides increases with increasing temperature and it has been observed that the phonon thermal conductivity dominates over total thermal conductivity in the entire temperature range studied. Phonon thermal conductivity can be further reduced by proper substitution or microstructure to enhance the figure of merit. The obtained maximum *ZT* value observed is 0.09 at 620 *K* for x=0.05. This suggests that slight addition of LSMO (x=0.05 in (1-x)$LaCoO_3$ .x$La_{0.7}Sr_{0.3}MnO_3$) not only enhances the value of *ZT* but also enhances the operating temperature of the TE materials.

## Acknowledgments

AK and ADT would like to acknowledge Ministry of Human Resources and Development (MHRD) India for funding during all the research work done.

**FIGURE CAPTIONS:**

**Fig.1.** (Color Online) X-ray diffraction pattern of (1-x)$LaCoO_3$.x$La_{0.7}Sr_{0.3}MnO_3$) composite. Bragg's position and ICDD file number are shown for LCO(black) and LSMO (green).

**Fig.2.** (Color online) Rietveld refinement pattern of (1-x)$LaCoO_3$.x$La_{0.7}Sr_{0.3}MnO_3$) composite sample for x=0.05.

**Fig.3.** (Color online) Field emission scanning electron micrograph of (1-x)$LaCoO_3$.x$La_{0.7}Sr_{0.3}MnO_3$) composite sample (a) x=0.00, (b-d) x=0.05

**Fig.4.** (Color online) (a) Seebeck coefficient (α) and (b) Electrical conductivity ($\sigma$) as a function of temperature for (1-x)$LaCoO_3$.x$La_{0.7}Sr_{0.3}MnO_3$) composite sample.

**Fig.5.** (Color online) Precision Seebeck measurement of (1-x)$LaCoO_3$.x$La_{0.7}Sr_{0.3}MnO_3$) composite sample (a) x=0.05 (b) x=0.10, (c) x=0.20 and (d) x=0.50. α values are measured in (μV/K).

**Fig. 6.** (Color Online) (a) Total thermal conductivity ($\kappa$), (b) electronic thermal conductivity ($\kappa_e$) and (c) phonon thermal conductivity ($\kappa_{ph}$) as a function of temperature measured using TPS for (1-x)$LaCoO_3$.x$La_{0.7}Sr_{0.3}MnO_3$) composite sample.

**Fig. 7**. (Color Online) Figure of merit (*ZT*) as a function of temperature calculated using thermoelectric parameters for (1-x)$LaCoO_3$.x$La_{0.7}Sr_{0.3}MnO_3$) composite sample.

**TABLE CAPTIONS:**

**TABLE I.** Refinement parameters $R_{wp}$ , $R_{exp}$ , $R_p$ , $\chi^2$ , Lattice parameters *a*, *c* and volume of the unit cell for (1-x)$LaCoO_3$ .x$La_{0.7}Sr_{0.3}MnO_3$) composite sample calculated from the Rietveld refinement of the XRD patterns.

**TABLE II.** Atomic position and their occupancy for $LaCoO_3$ and $La_{0.7}Sr_{0.3}MnO_3$ polycrystalline sample.

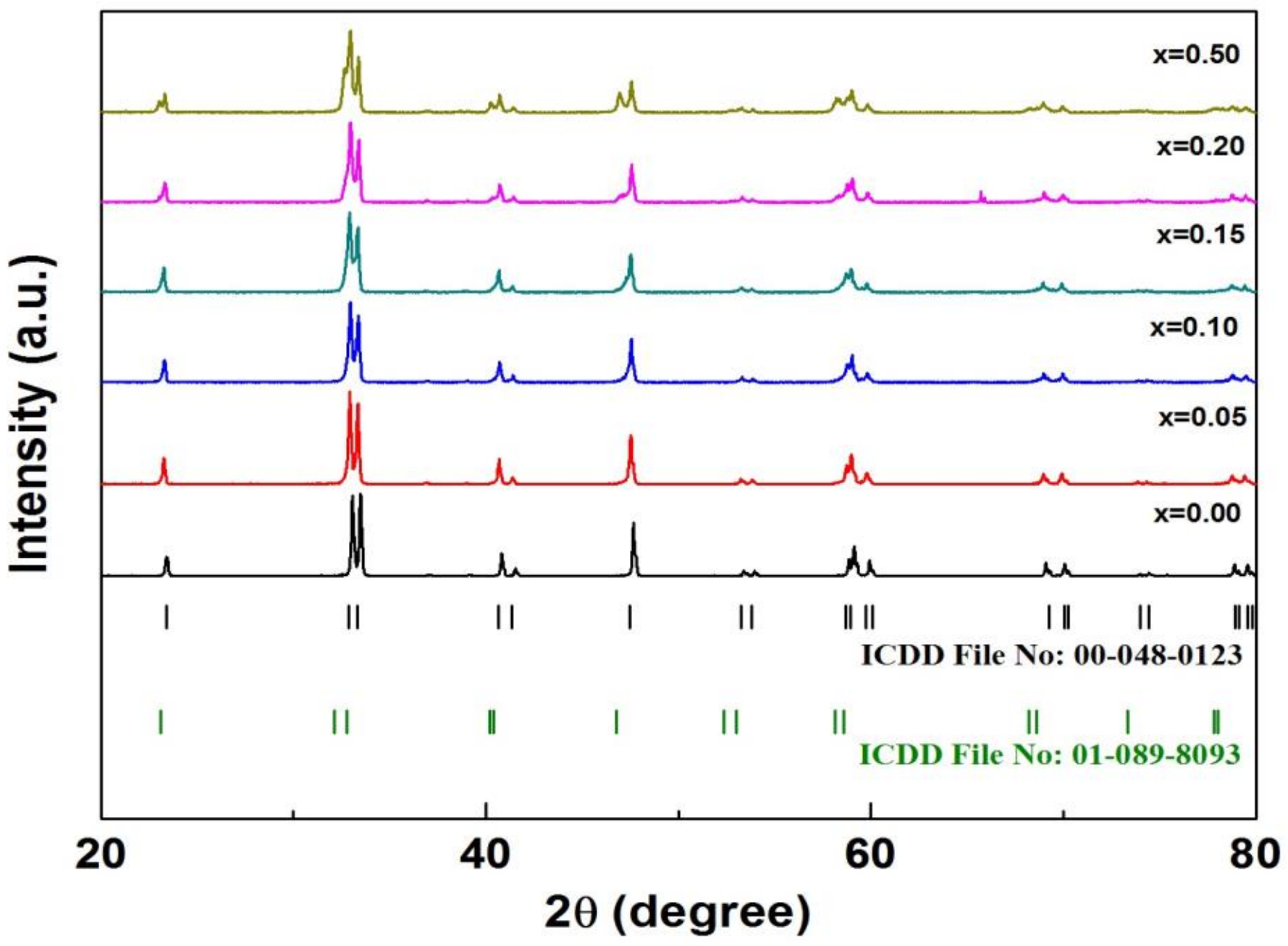

**Fig.1.** (Color Online) X-ray diffraction pattern of (1-x)$LaCoO_3$ .x$La_{0.7}Sr_{0.3}MnO_3$) composite. Bragg's position and ICDD file number is shown for LCO (black) and LSMO (green).

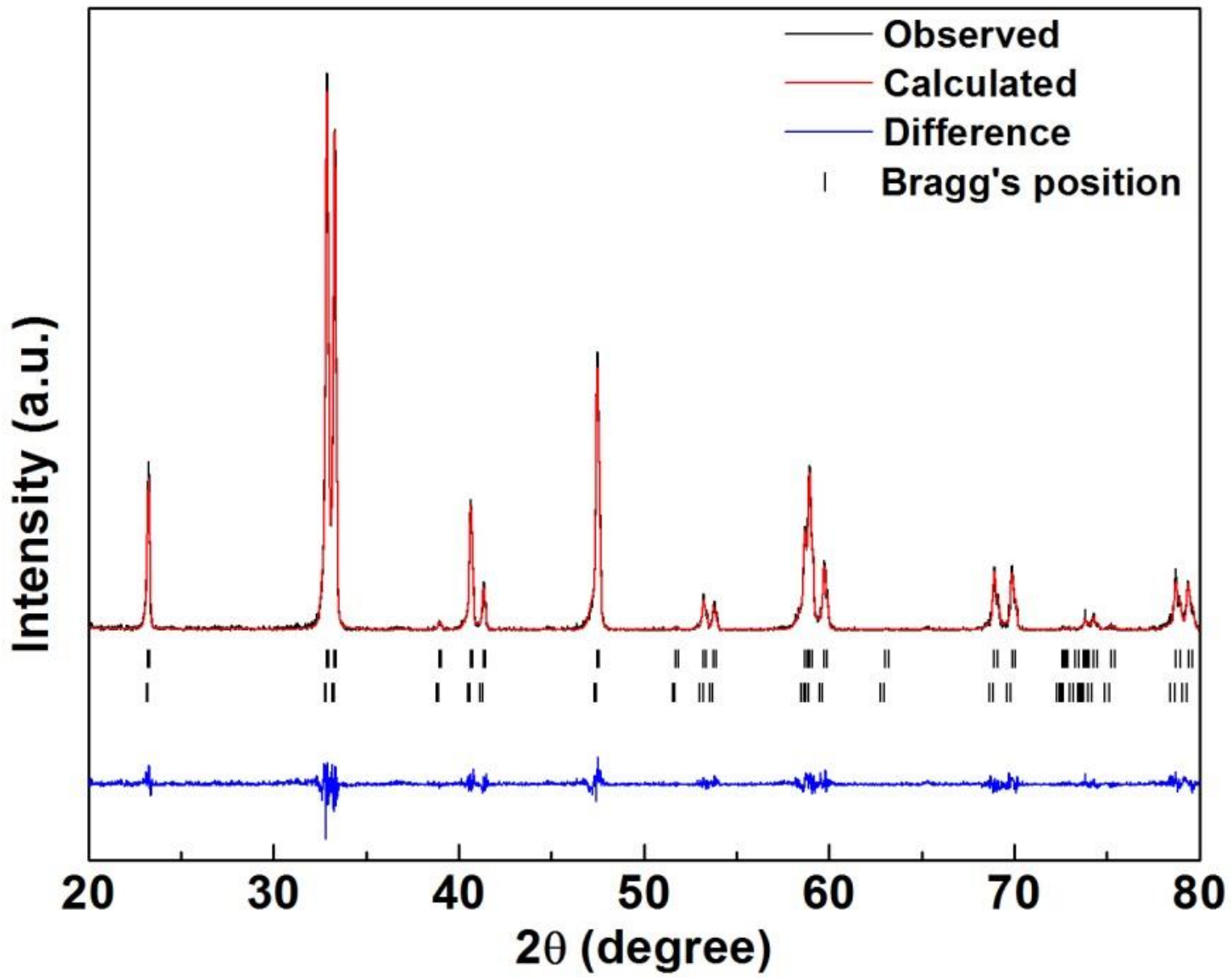

**Fig.2.** (Color online) Rietveld refinement pattern of (1-x)$LaCoO_3$.x$La_{0.7}Sr_{0.3}MnO_3$) composite sample for x=0.05.

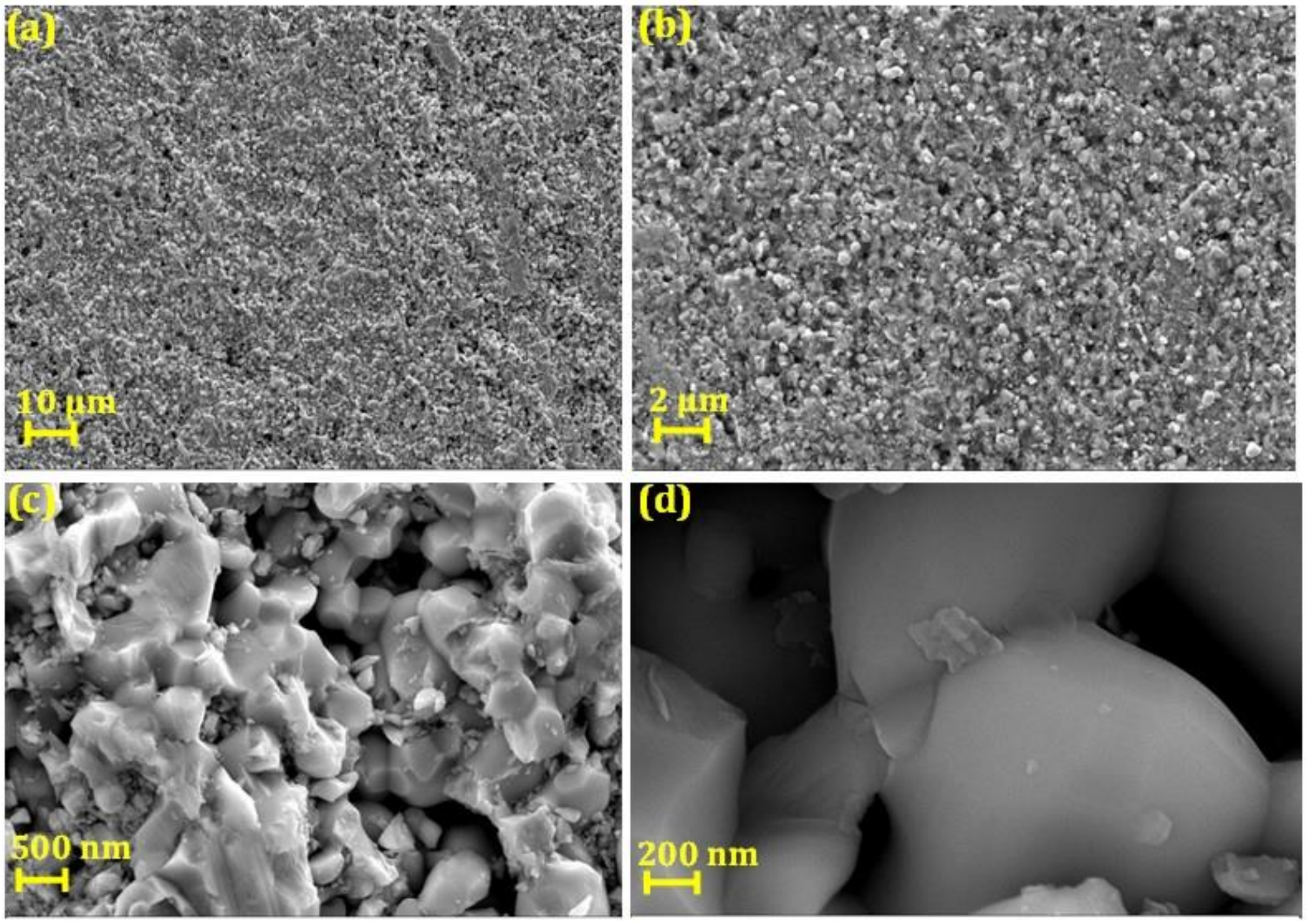

**Fig.3.** (Color online) Field emission scanning electron micrograph of (1-x)$LaCoO_3$.x$La_{0.7}Sr_{0.3}MnO_3$) composite sample (a) x=0.00, (b-d) x=0.05

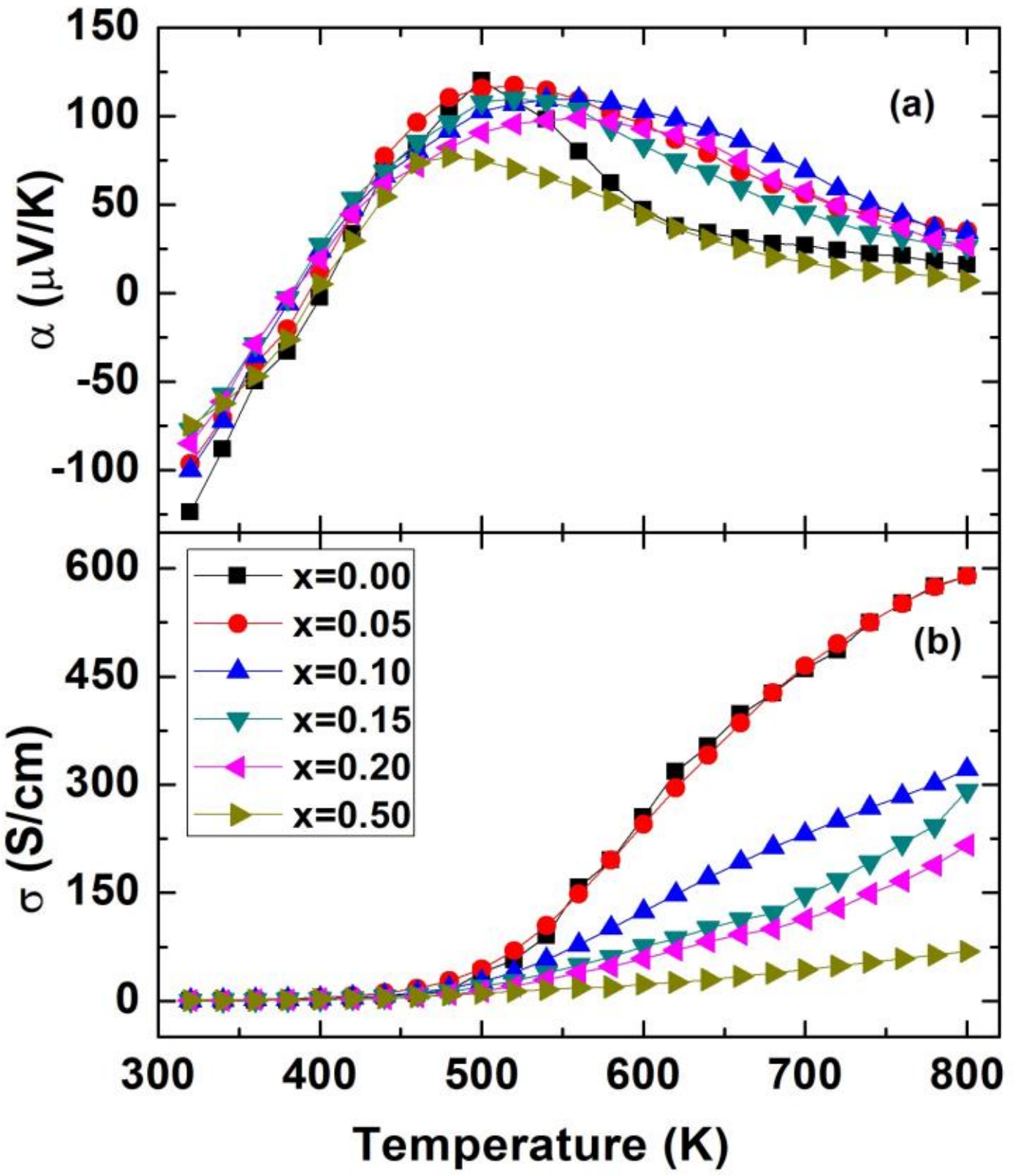

**Fig.4.** (Color online) (a) Seebeck coefficient (α) and (b) Electrical conductivity ($\sigma$) as a function of temperature for (1-x)$LaCoO_3$.x$La_{0.7}Sr_{0.3}MnO_3$) composite sample.

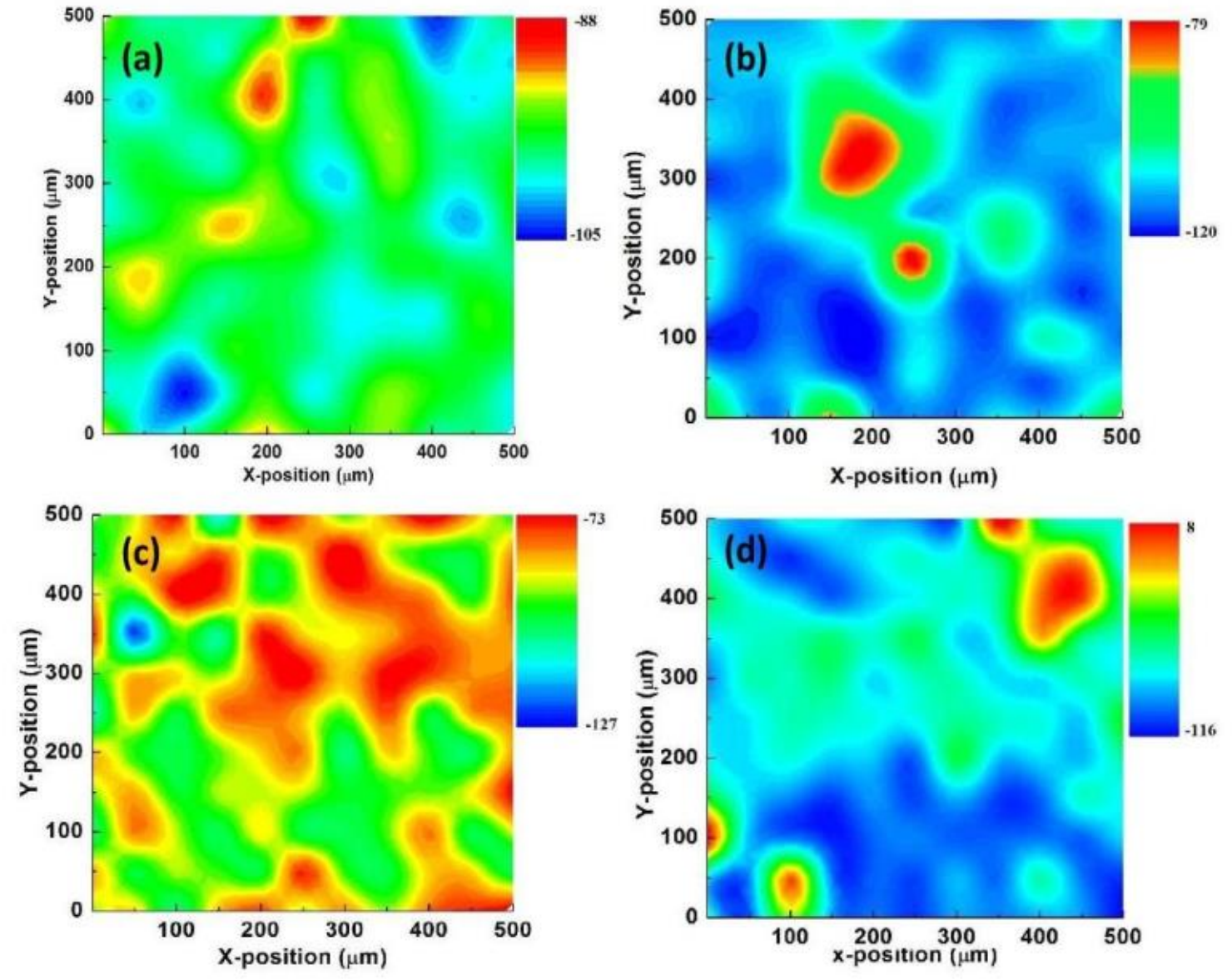

**Fig.5.** (Color online) Precision Seebeck measurement of (1-x)$LaCoO_3$.x$La_{0.7}Sr_{0.3}MnO_3$) composite sample (a) x=0.05 (b) x=0.10, (c) x=0.20 and (d) x=0.50. α values are measured in (μV/K).

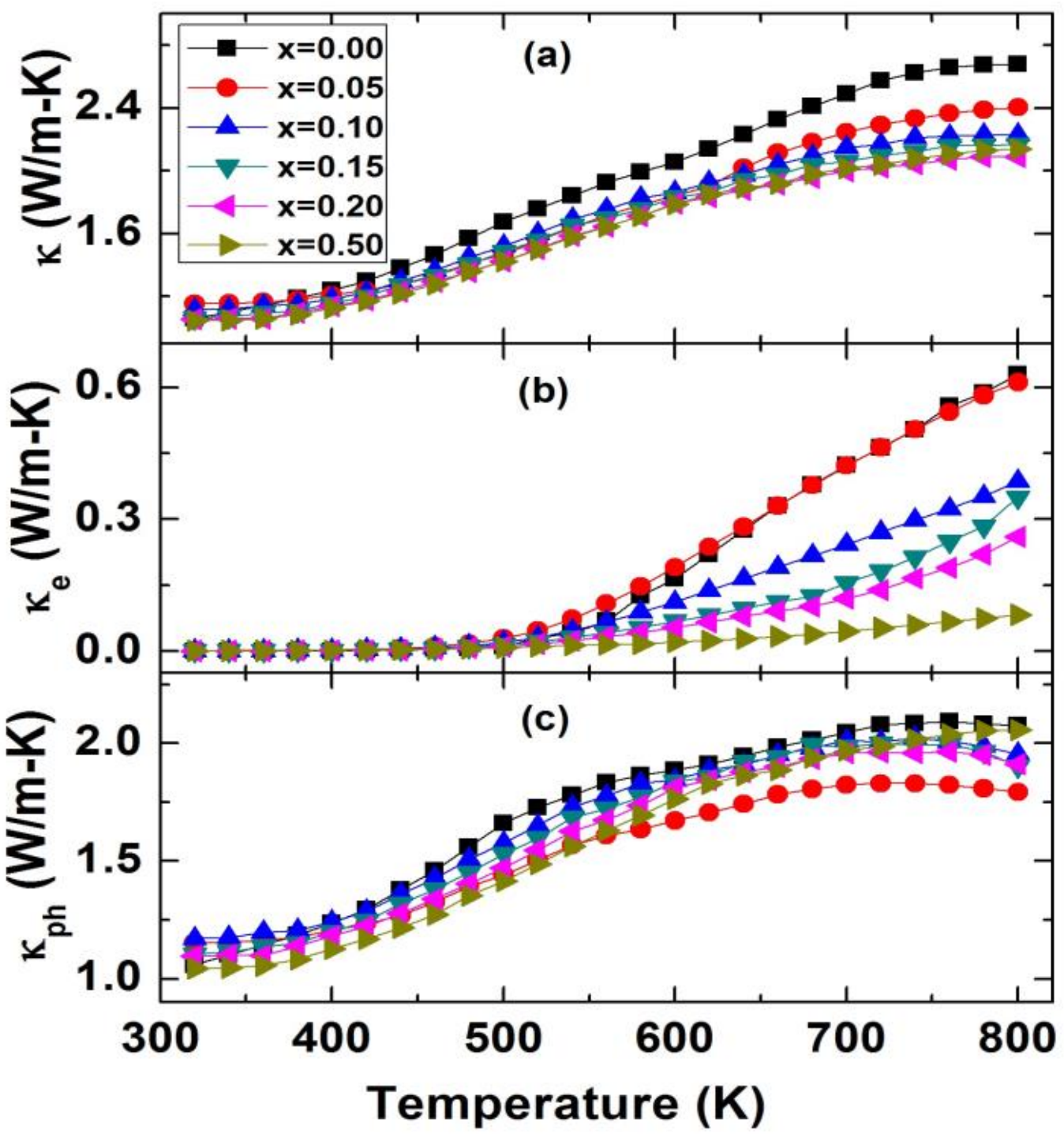

**Fig. 6.** (Color Online) (a) Total thermal conductivity (*κ*), (b) electronic thermal conductivity ($\kappa_e$) and (c) phonon thermal conductivity ($\kappa_{ph}$) as a function of temperature measured using TPS for (1-x)$LaCoO_3$.x$La_{0.7}Sr_{0.3}MnO_3$) composite sample.

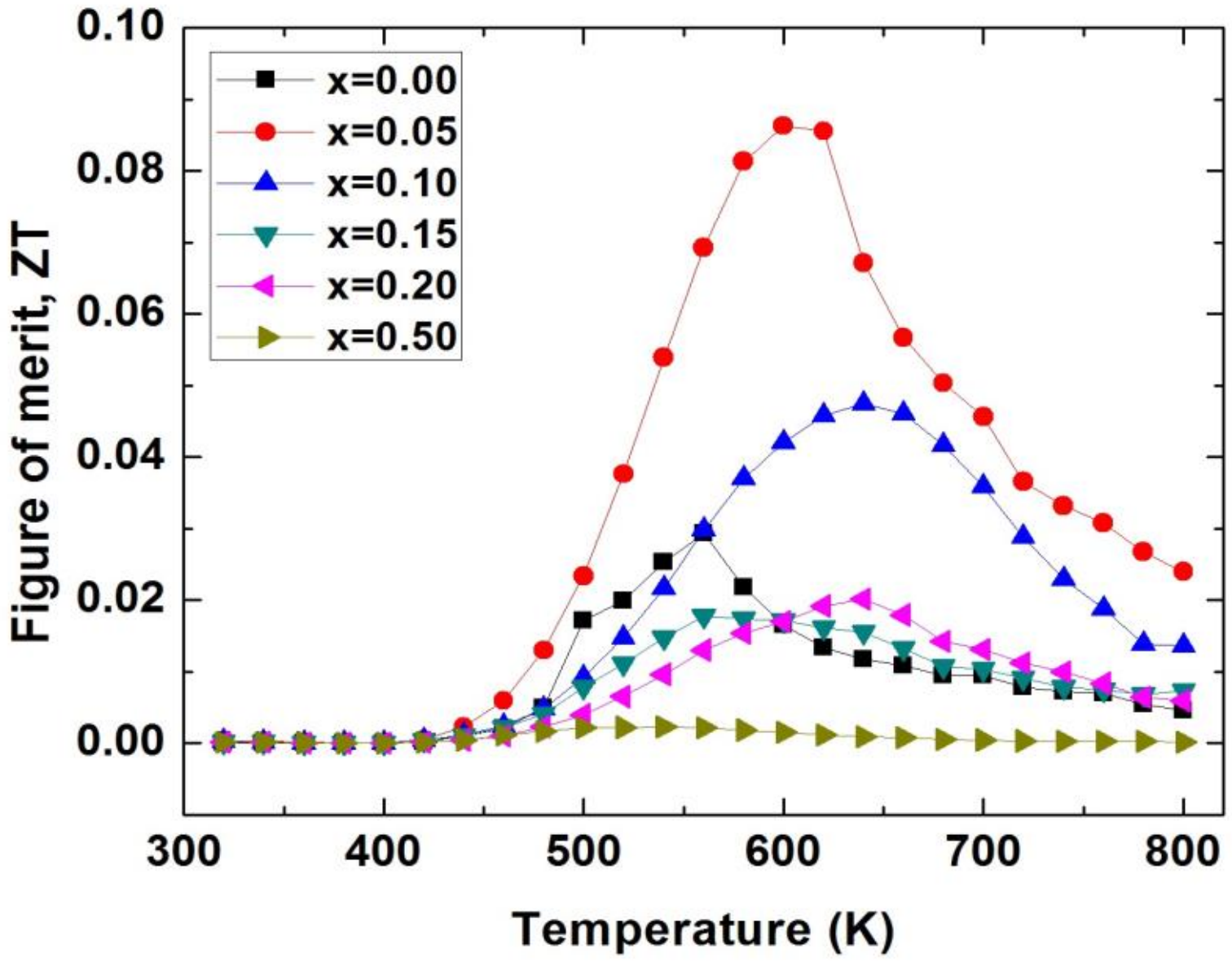

**Fig. 7**. (Color Online) Figure of merit (*ZT*) as a function of temperature calculated using thermoelectric parameters for (1-x)$LaCoO_3$.x$La_{0.7}Sr_{0.3}MnO_3$) composite sample.

**TABLE I.** Refinement parameters $R_{wp}$ , $R_{exp}$ , $R_p$ , $\chi^2$ , Lattice parameters *a, c* and volume of unit cell for (1-x)$LaCoO_3$.x$La_{0.7}Sr_{0.3}MnO_3$) composite sample calculated from the Rietveld refinement of the XRD patterns.

| **$LaCoO_3$ (1-x)-$La_{0.7}Sr_{0.3}MnO_3$(x)** | | **a (Å)** | ***c* (Å)** | **V(Å$^3$)** | $R_{wp}$ | $R_{exp}$ | $R_p$ | $\chi^2$ | **Phase %age** |
|---|---|---|---|---|---|---|---|---|---|
| **x=0.00** | $LaCoO_3$ | 5.439(5) | 13.084(6) | 335.28 | 23.3 | 18.19 | 19.3 | 1.64 | 100 |
| **x=0.05** | $LaCoO_3$ | 5.445(1) | 13.099(6) | 336.35 | 23.7 | 17.30 | 14.5 | 1.87 | 94.2 |
| | $La_{0.7}Sr_{0.3}MnO_3$ | 5.462(2) | 13.147(2) | 339.71 | | | | | 5.8 |
| **X=0.10** | $LaCoO_3$ | 5.444(3) | 13.094(6) | 336.13 | 22.7 | 16.80 | 13.4 | 1.84 | 89.7 |
| | $La_{0.7}Sr_{0.3}MnO_3$ | 5.461(6) | 13.151(4) | 339.73 | | | | | 10.3 |
| **x=0.15** | $LaCoO_3$ | 5.445(4) | 13.094(6) | 336.26 | 22.8 | 16.4 | 13.2 | 1.94 | 84.2 |
| | $La_{0.7}Sr_{0.3}MnO_3$ | 5.465(0) | 13.161(5) | 340.42 | | | | | 15.8 |
| **x=0.20** | $LaCoO_3$ | 5.446(1) | 13.096(8) | 336.41 | 23.4 | 17.2 | 13.6 | 1.84 | 81.0 |
| | $La_{0.7}Sr_{0.3}MnO_3$ | 5.465(3) | 13.161(4) | 340.41 | | | | | 19.0 |
| **x=0.50** | $LaCoO_3$ | 5.446(5) | 13.109(4) | 336.78 | 22.8 | 16.4 | 13.3 | 1.93 | 51.2 |
| | $La_{0.7}Sr_{0.3}MnO_3$ | 5.477(8) | 13.202(4) | 343.07 | | | | | 48.8 |

**TABLE II.** Atomic position and their occupancy for $LaCoO_3$ and $La_{0.7}Sr_{0.3}MnO_3$ polycrystalline sample.

| | **Site** | **x** | **y** | **z** | **$B_{iso}$** | **Occupancy** |
|---|---|---|---|---|---|---|
| **$LaCoO_3$ ($R\bar{3}c$)** | | | | | | |
| **La** | **6a** | **0.000** | **0.000** | **0.250** | **0.00** | **0.1655** |
| **Co** | **6b** | **0.000** | **0.000** | **0.000** | **0.00** | **0.1855** |
| **O** | **18c** | **0.453** | **0.000** | **0.250** | **0.00** | **0.5126** |
| **$La_{0.7}Sr_{0.3}MnO_3$ ($R\bar{3}c$)** | | | | | | |
| **La/Sr** | **6a** | **0.000** | **0.000** | **0.250** | **0.00** | **0.1155/0.0503** |
| **Mn** | **6b** | **0.000** | **0.000** | **0.000** | **0.00** | **0.1662** |
| **O** | **18c** | **0.437** | **0.000** | **0.250** | **0.00** | **0.4605** |